\documentclass[11pt]{article}
\RequirePackage[OT1]{fontenc}
\usepackage{amsthm,amsmath,amssymb,amsfonts,bm,enumerate,xcolor,graphicx,natbib,color,url,graphbox}
\RequirePackage[colorlinks,citecolor=blue,urlcolor=blue]{hyperref}
\usepackage{hyperref}
\usepackage{subcaption} 
\usepackage{ulem}
\usepackage{pifont}
\usepackage{textcmds}
\usepackage{booktabs}
\usepackage{multirow}
\usepackage{booktabs}
\usepackage{adjustbox}
\usepackage[usestackEOL]{stackengine}
\usepackage[toc,page]{appendix} 
\usepackage{algorithm,algorithmic}
\usepackage{lipsum}
\usepackage{setspace}

\usepackage{setspace}
\allowdisplaybreaks
\usepackage{makeidx}
\usepackage{multicol}

\newcommand{\blind}{1}

\begin{document}
%%%%%%%%%%%%%%%%%%%%%%%%%%%%%%%%%%%%%%%%%%%%%%%%%%%%%%%%%%%%%%%%%%%%%%%%%%%%%%
\thispagestyle{empty}
\baselineskip=28pt
\vskip 5mm

\renewcommand{\thefootnote}{\fnsymbol{footnote}}

\begin{center} 
{\Large{\bf Statistics of Extremes for Neuroscience}}\footnotemark[2]
\end{center}

\baselineskip=12pt

\vskip 5mm

\renewcommand{\thefootnote}{\arabic{footnote}}

\if1\blind
{
\begin{center}
\large
Paolo Redondo$^1$, Matheus B. Guerrero$^2$, Rapha\"el Huser$^1$, and Hernando Ombao$^1$
\end{center}
\renewcommand{\thefootnote}{\fnsymbol{footnote}} \footnotetext[2]{Extract of Chapter 30 (first 5 pages), to be published in the \emph{Handbook of Statistics of Extremes}, eds.\ Miguel de Carvalho, Rapha\"el Huser, Philippe Naveau and Brian Reich}
\renewcommand{\thefootnote}{\arabic{footnote}} \footnotetext[1]{
\baselineskip=10pt Statistics Program, Computer, Electrical and Mathematical Sciences and Engineering (CEMSE) Division, King Abdullah University of Science and Technology (KAUST), Thuwal 23955-6900, Saudi Arabia. E-mails: paolovictor.redondo@kaust.edu.sa, raphael.huser@kaust.edu.sa, hernando.ombao@kaust.edu.sa}
\footnotetext[2]{
\baselineskip=10pt  Department of Mathematics, California State University, Fullerton, California. E-mail: matguerrero@fullerton.edu}
\fi

\baselineskip=26pt
\vskip 2mm
\centerline{\today}
\vskip 4mm

%%%%%%%%%%%%%%%%%%%%%%%%%%%%%%%%%%%%%%%%%%%%%%%%%%%%%%%%%%%%%%%%%%%%%%%%

\baselineskip=14pt

This chapter illustrates how tools from univariate and multivariate statistics of extremes can complement classical methods used to study brain signals and enhance the understanding of brain activity and connectivity during specific cognitive tasks or abnormal episodes, such as an epileptic seizure.

\section{Introduction}\label{sec:intro}

Understanding the dynamics of how the brain works is one of the primary goals in neuroscience. Given various modalities for collecting brain data, e.g., functional magnetic resonance imaging (fMRI) and electroencephalography (EEG), a wide range of research has led to many interesting findings and practical developments in the medical community. Some applications involve predicting epileptic seizure onset \citep{alotaiby2014eeg,boonyakitanont2020review}, classification of patients with neurophysiological disorders from healthy subjects \citep{mohammadi2016eeg,ieracitano2020novel} and advancements on the so-called brain-computer interfaces (BCI), machines that decipher brain activity and translate them to physical actions such as controlling robotic arms for people with paralysis \citep{ang2015randomized,gillini2021assistive,altaheri2023deep}. Yet, despite having evidence on the heavy-tailed nature of the human brain \citep{freyer09,roberts15} and the fact that large amplitudes of brain electrical signals are often associated with cognitive activity or disruptions \citep{macneilage1966eeg,brismar2007human,sauseng2008does}, most of the current methods for analyzing brain data deal with inference on central trends, focusing on describing the bulk of the data distribution. Such practice neglects the tail behavior of the complex mechanisms of brain physiological processes, which may arguably provide more informative scientific insights, especially in cases of alterations in brain functional activity due to some mental illness.

Although limited so far, several papers in the literature have used extreme value theory (EVT) in the analysis of brain data. For applications in MRI data, \cite{roberts2000extreme} developed novelty detection using EVT, while \cite{dawkins2019theoretical} incorporated EVT in the derivation of the theoretical properties of nearest-neighbor distance distributions. In EEG data analysis, \cite{karpov2022extreme} proposed machine learning approaches for seizure detection that are inspired by EVT, and \cite{vrba2020introduction} developed the extreme seeking entropy algorithm for novelty detection. There are also applications of EVT to epilepsy data \citep{luca2014detecting,pisarchik2018extreme,frolov2019statistical}. However, the commonality among these works is the use of extreme value distributions to fit the margins of the data, which simply serves as a preliminary step prior to the main methodology (e.g., deep learning algorithm). To our knowledge, the very first paper in brain data analysis that is ``fully'' EVT-based is \cite{guerrero23}, who implemented the conditional extremes model of \cite{HT04} to investigate the behavior of the signals from a channel given that a reference channel exhibits extreme amplitudes. We review and apply their methodology in Section~3.3. However, the application of EVT in neuroscience still requires many developments and further evidence of its practical usefulness to make an impact in the field and become widely spread.

The goal of this chapter is to demonstrate how models and techniques in EVT can be utilized to analyze brain signals, specifically to EEG data, and show that EVT can provide useful insights into brain abnormalities, e.g., during an epileptic seizure.

\subsection{Background}

Electroencephalograms (EEGs) are recordings of electrical signals generated during brain activity that are measured from the cortical surface of the head \citep{binnie1994electroencephalography,ombao2016handbook}. In the recent years, EEG is considered as one of the leading clinical tools for investigating neurological diseases such as epilepsy \citep{noachtar2009role,benbadis2020role}, schizophrenia \citep{o2006role,haenschel11} and dementia \citep{adamis2005utility,al2014role,law2020role},
because it is non-invasive, easy to collect, relatively inexpensive, and accommodates high temporal resolution (up to 1000 Hertz/samples per second) \citep{lenartowicz2014use}. However, since EEGs are measured by placing electrodes on the scalp (see Figure~\ref{fig:eeg_freqbands}), one limitation is that it has a lower resolution than fMRI data, and that it is highly sensitive to noise artifacts (which includes muscle movements, eye blinking and electrical surges from devices or the EEG machine itself). Thus, prior to exploratory analysis and statistical modeling of EEG data, it is a standard practice to remove such artifacts by applying pre-processing pipelines to improve the quality of the signals \citep{kim2018preprocessing}.

\begin{figure}[t!]
    \centering
    \includegraphics[width = 0.45\textwidth]{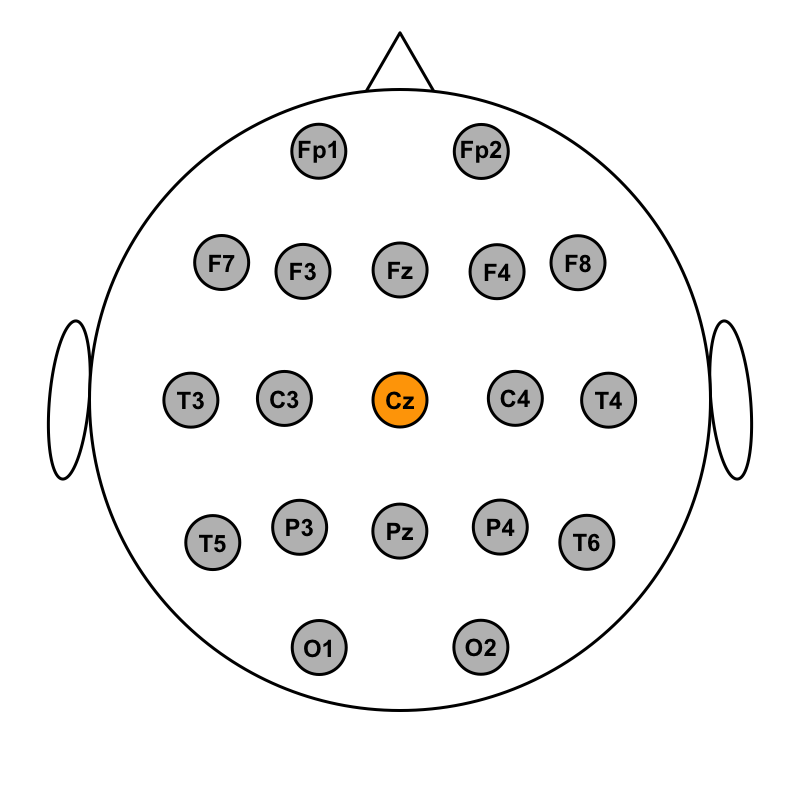}
    \includegraphics[width = 0.45\textwidth]{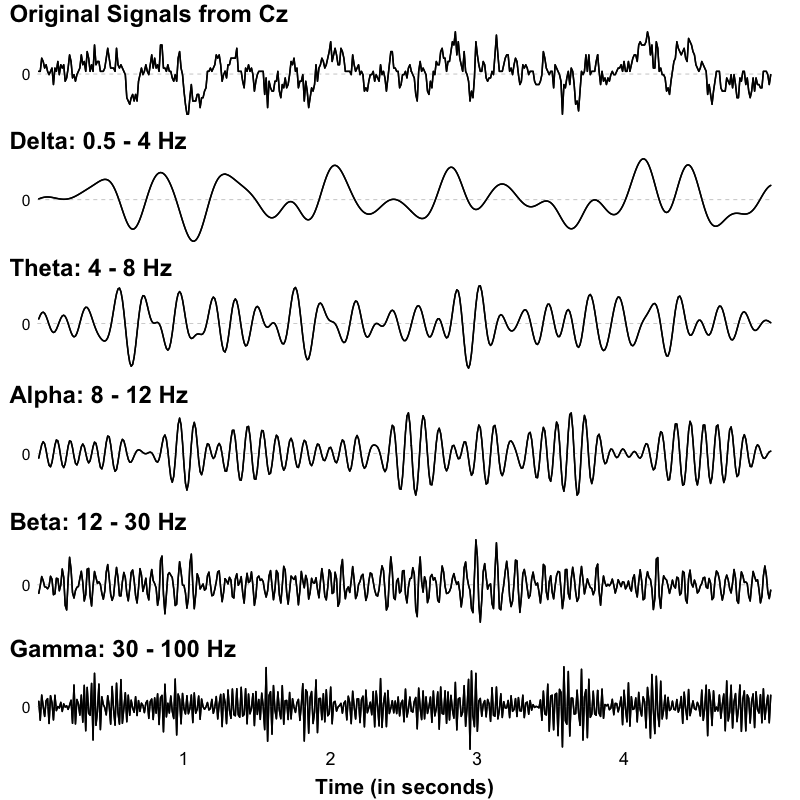}
    \caption{Standard 10--20 EEG scalp topography with 19 channels and decomposition of EEG recordings into the five standard frequency bands.}
    \label{fig:eeg_freqbands}
\end{figure}

In addition, EEGs are often characterized as a superposition of random oscillations at various frequencies, from slowly oscillating waveforms (low frequency waves) to rapidly oscillating waveforms (high frequency waves). In practice in neurology and cognitive neuroscience, EEGs are thought to be composed of different frequency oscillations grouped into five standard waveforms/frequency bands commonly called --- delta ($\Omega_1$: 0.5–4 Hz), theta ($\Omega_2$: 4–8 Hz), alpha ($\Omega_3$: 8–12 Hz), beta ($\Omega_4$: 13–30 Hz), and gamma ($\Omega_5$: 30–100 Hz) \citep{cohen2017does,gao2020,ombao2022spectral,granados2024bayesian}. There are numerous evidences that link these band-specific oscillatory processes to various cognitive functions. For example, low frequency waveforms are associated with sleep and attention while high frequency oscillations are related to complex neuronal activation and performance of motor movements \citep{herrmann2016eeg}. This is the main reason why most analysis of EEG data rely on spectral decomposition of the signals into the standard frequency bands through linear filtering (e.g., Butterworth and finite impulse response filter) or use a wavelet-based approach \citep{nunez2016electroencephalography,fiecas2016modeling,ombao2016handbook}. EVT's role is to offer additional tools that complement existing methodologies in providing insights to understanding brain dynamics.

\subsection{Motivating Dataset}\label{sec:ep_data}

In this chapter, we illustrate how extreme behavior in EEG data can be rigorously analyzed through EVT, using a well-studied dataset \citep{ombao05,schroder2019fresped,guerrero23}. We consider the EEG recordings are collected from a female patient diagnosed with left temporal lobe epilepsy whose focal point of seizure often occurs in the temporal area of the left hemisphere of the brain, closest to the reference channel T3 (see Figure~\ref{fig:epilepsy_data}). This EEG dataset contains $500$ seconds of recording at a sampling rate of $100$ Hz (yielding $T =$ 50,000 time points). For this dataset, the attending neurologist noted that the seizure onset happens roughly at the $350^{th}$  second (indexed at time point $t = $ 35,000). 

\begin{figure}[t!]
    \centering
    \includegraphics[align=c,width = 0.25\linewidth]{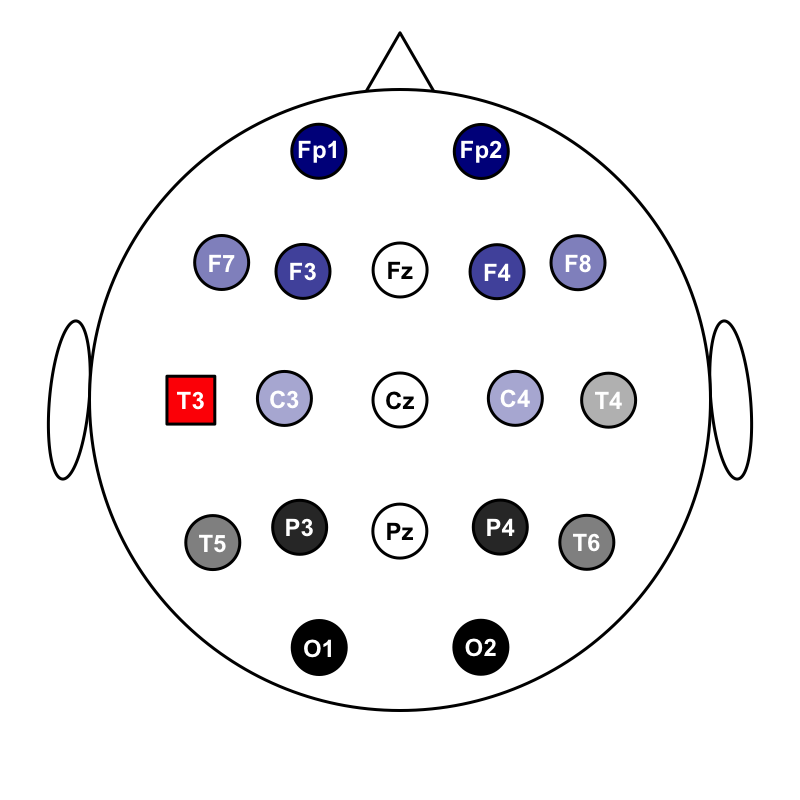}
    \includegraphics[align=c,width = 0.65\linewidth]{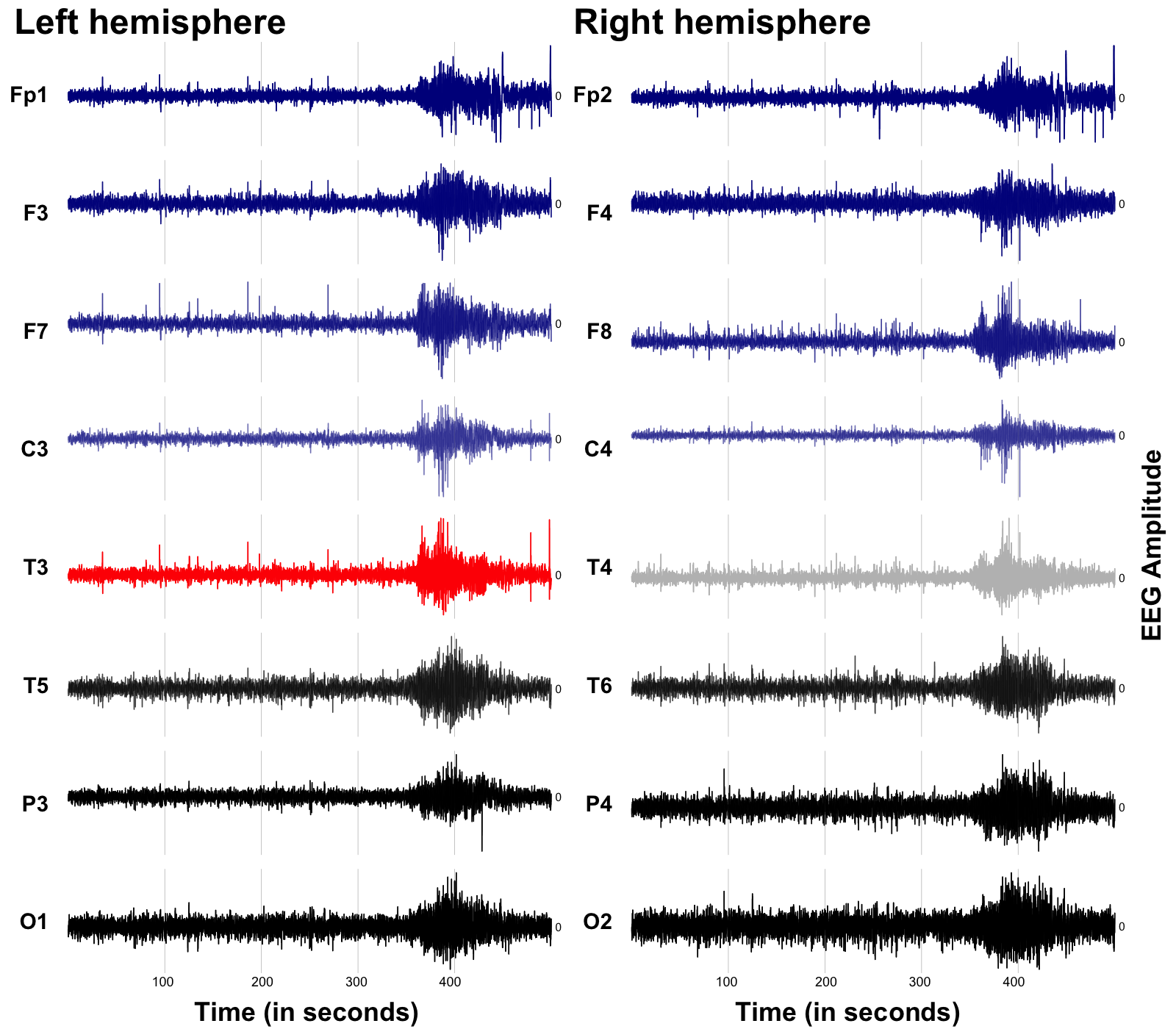}
    \caption{Recorded EEG signals from selected channels during an epileptic seizure episode of a female patient diagnosed with left temporal lobe epilepsy.}
    \label{fig:epilepsy_data}
\end{figure}

Clearly, the behavior of the EEG signals before the onset of epileptic seizure differs from the behavior after the onset. One can observe the excessively large (extreme) amplitudes exhibited by all channels during the seizure episode. This is a natural example of an extreme event in brain signals which motivates the use of EVT in analyzing such data.

\subsection{Chapter organization}

The rest of the chapter is organized as follows. An overview of common methodologies for EEG analysis in the frequency domain is presented in Section~2. In Section~3, we illustrate how univariate extremes modelling of EEG data is performed, and demonstrate how extremal dependence may be quantified using empirical diagnostics and through the conditional extremes model, thus complementing existing techniques. Lastly, a brief summary and avenues for future research on the application of EVT in neuroscience are discussed in Section~4.

\baselineskip 12pt
\bibliographystyle{CUP}
\bibliography{bibliography}

\end{document}